# Blended e-Learning Training (BeLT): Enhancing Railway Station Controller Knowledge

Aditya Khamparia
Department of Computer Science and Engineering,
Lovely Professional University,
Phagwara, Punjab, India
aditya.khamparia88@gmail.com

Monika Rani
Department of Computer Science,
Indian Institute of Information Technology,
Allahabad, Uttar Pradesh, India
monikarani1988@gmail.com

Babita Pandey
Department of Computer Applications,
Lovely Professional University Phagwara,
Punjab, India
shukla_babit@yahoo.com

O. P. Vyas
Department of Computer Science,
Indian Institute of Information Technology,
Allahabad, Uttra Pradesh, India
dropvyas@gmail.com

*Abstract*— With the growing economy, e-learning consequently gained increasing attention as it conveys knowledge globally with improved interactivity, assistance and reduced costs. For the past few years, accidental challenges have become the severe problem with railway units due to irresponsibility, lack of knowledge and improper guidance of station controllers (learners). While focusing on e-learning technologies railway units failed to admit learner's need, cultural diversity and background skills by creating ethnically impartial e-learning environments, which resulted in inadequate training and degraded performance. The purpose of this study is to understand the vision of a global diverse group of station traffic controllers about e-learning courses developed by their individual railway units. The opinions of these officials have been verified by questionnaires on the basis of course organization, course accuracy, course effectiveness, course relevance, course productivity and course interactivity. The results obtained show that the developed e-learning course was highly helpful, interactive, creative, and user-friendly for learners. This leads to making e-learning conquered among independent learners.

## 1. INTRODUCTION

Due to enhancement in railway accidents, various officials of top level management decided to build up the professional growth of their station traffic controller employees by providing them super excellent training which leads to avoid damage. The training must provide a consistent learning based instruction with minimal hardware cost requirement. Railway units of different countries provide professional development of employee through e-learning which enriched the growing magnetism of this method of online training [1]. These methods surpassed the traditional or conventional classroom mode of training where the training not conducted on the basis of learners preference, background, skills, etc. due to which learner not able to get rid of risk based situations. The blend e-learning [2] teaching approach is useful to enhance the knowledge of the station controllers (learners) for the railway station training program. Blended e-learning training (BeLT) drastically improves the culture of western and non-western economy, which relied mostly on the development of theoretical valued instructions rather than their application usages in diverse environments. According to Hannon et al. usage of learning technologies negatively influenced the level of engagement for culturally manifold learners [3]. The purpose of this research is to study the vision of a global diverse group of station traffic controllers about e-learning courses which was developed by their countries railway units. Based on relevant literature, there were total 30 null hypotheses clumped into six groups according to defined form as: course relevance, course organization, course effectiveness, course interactivity, course accuracy and course productivity. The various types of null hypotheses were: (1) there was no significant difference in course relevance in impartial e-learning environment for defendants country of origin, cultural diversity, religion and philosophy, arts & music and native languages; (2) there was insignificant difference in course organization for defendants country of origin, cultural diversity, religion and philosophy, arts & music and native languages; (3) there was insignificant difference in course effectiveness for defendants country of origin, cultural diversity, religion and philosophy, arts & music and native languages; (4) there was insignificant difference in course interactivity for defendants country of origin, cultural diversity, religion and philosophy, arts & music and native languages; (5) there was insignificant difference in course accuracy for defendants country of origin, cultural diversity, religion and philosophy, arts & music and native languages; (6) there was insignificant difference in course productivity for defendants country of origin, cultural diversity, religion and philosophy, arts & music and native languages.

## 2. LITERATURE REVIEW

In the last few years, railway units of different countries faced numerous challenges in terms of accidents, delay in arrival and departure of trains, location detection and financial downturn that devastated the economy badly. As to manage the traffic and track detection, station traffic controller officials played a crucial role which became the supreme authority to provide excellent customer services for railway industries. Railway units developed an improvement strategy for achieving operational and environmental excellence with the help of traffic controllers and competitiveness of railway markets, which was monitored by customer satisfaction surveys [4] [5].

To achieve such requirements, railway units hired special trained traffic controller officials and provided training to them. Recruitment of officials should be such that it enhanced the growth of railway unit and their productivity. After successful recruitments, training of individual controllers also important, according to job position through various learning based models like instructional system design, Simulink etc. The various processes like data analysis, instruction planning, efficient implementation of instruction and regular performance evaluation of training products through the years [6].

Various trainings conducted in terms of e-learning, online video monitoring, surveillance based for development of interpersonal skills of station controllers. E-learning techniques [7] [8] had various advantages over conventional classroom techniques including cost efficiency (savings); just in time access to timely information; higher retention of content; higher interactivity; improved collaboration, etc. Disadvantages of e-learning included lack of social skill development; learners inability to use computers. E-learning system were based on the assumption culture, values were involved in the reconcile difference of use updated technology and collaborate instructions in order to conserve culture environment. McLoughlin et al. allow student centering learning by providing an opportunity to adapt culture based learning according to skills, beliefs and styles of the student [9]. A1-Hunaiyyan et al. mention language as a constraint for educational software as it is issued for users who doesn't understand the English language as instruction material general are in English language, it's hard to provide instruction in the native language[10]. Passarella et al., had designed a train obstacle detection system in which infra-red rays were used for identification of collision [11]. The ideal learning environment should consider the multicultural model of instructional design which consist of multicultural realities of society, various ways of teaching and learning and its outcomes [12]. Sambhamurty et al. proposed a wireless based network for prevention of train collision and its avoidance with disrespect to cultural organization, arts and music, productivity and accuracy etc. [13]. Miwa et al. developed statistical analysis based strategy to eliminate the accidental cause and effective step to avoid train accidents in Japan [14].Yaghini et al. use neural network model with high accuracy for the prediction the delay of passenger train [15].The design and development of training in global environment require understanding of various aspects like organization, of course, course productivity, accuracy, diversity, course effectiveness and its interactivity [16]. These ideas had been captured by railway units and tried to train station controllers to avoid various impartiality during their training and test conductance which leads to improvise learning [17, 18].

## 3. METHODOLOGY

The comparative methodology is used to explore the study which utilizes different questionnaires to construct and endorse the cultural connection in course designing and e-learning efficacy in global environments. Different questions of MCQ type were developed for station controllers who had taken participation in e-learning courses during their job joining at different railway headquarters. Data based on population growth in native regions, cultural diversity, religion and philosophy, music & arts, linguistic language or native languages were collected and used as an independent variable. The interaction between learner and program driven instructions, program organization and different multimedia contents assessed the contingent builds like Course organization, Course effectiveness, Course interactivity, Course accuracy, Course relevance and course productivity.

A questionnaire was sent to station controllers who were taking part in e-learning courses and they need to submit their response within stipulated time. A Web based program called SurveyGizmo was utilized for conveyance and feedback or response collection. For data analysis the Statistical Package for Social Sciences (SPSS) was deployed. To examine the station controllers a pre-test was conducted to evaluate the questionnaire credibility and its calibration with proposed hypothesis. To measure the internal consistency of questionnaire, i.e. how group of seven questions is interrelated, a Cronbach's alpha reliability analysis was used. Its value for different set of questions as: course organization ($\alpha=0.765$), effectiveness of e-learning course ($\alpha=0.722$), interactivity and navigation of course ($\alpha=0.788$), course accuracy ($\alpha=0.612$), course relevance ($\alpha=0.522$) and productivity of e-learning courses ($\alpha = 0.698$).

In this proposed study, we have considered Asia's geographic regions which are classified into different groups as: India (South Asia), Japan (East Asia), Iran (West Asia), Taiwan (East Asia) and Malaysia (Southeast Asia). Based on defendant's country of origin, their cultural diversity was classified as: western or non-western. Religion and Philosophy classified majorly as: Christian or non-christian. Similarly, arts and music were classified as: classical or non-classical. Finally the native languages of the defendants were classified as: English speaker or non-English speaker. Different defendant's response on e-learning environments was captured and analyzed. Likert scale was used to calculate the response and numerical values were attributed to each scale parameter: "Totally agree" as 1, "Agree" as 2, "Neutral" as 3, "Disagree" as 4 and "Totally disagree" was assigned 5.

To match the literary factors with statistical factors stable technique called factor analysis has been used. Various factors were used in the testing of hypothesis. Different hypothesis on demographic data about origin were tested using ANOVA technique and rest other hypothesis were identified or tested using t or f test.

## 4. RESULTS

The contemplate was sent via mailing mode to 100 station controllers working in the different railway headquarters unit in Asia region. Out of 100 station controllers only 45 candidates able to accessed the questionnaire which leads to an access rate of 45%. Out of these 45 defendants, only 32 able to fully complete the survey and created response rate of 32%. The defendants were from India (31%), Japan (22%), Iran (19%), Taiwan (17%) and Malaysia (11%). Out of these, 42% were English speakers and 58% were non-English speakers. Similarly, 61% were Christian and 39% were non-Christian. For art and music, 48% were emphasized on classical music and 52% were on non-Classical and finally 42% were categorized as Western and 58% were categorized as non-Western. Different types of questionnaire review were conducted on the basis of factor analysis on different relevant groups of survey questions. Data statistics suitability for factor analysis was identified on the basis of Kaiser-Meyer-Olkin measure of sampling adequacy and Bartlett's test of Sphericity. Various types of formulations were created by factor analysis as: (1) Course organization, (2) Course effectiveness, (3) Course interactivity, (4) Course accuracy, (5) Course relevance and (6) Course productivity. Various formulations were tested hypotheses with the five independent variables as shown in Table 1, 2, 3, 4, 5 and 6. After getting the results of hypothesis testing on different formulation, researchers concluded that defendants had a diverse belief about impartial e-learning. Different countries of Asia are denoted as In, India; Ja, Japan; Ir, Iran; Ta, Taiwan and Ma, Malaysia. Religion denoted as Ch, Christian; N-Ch, non-Christian. Music and art denoted as Cu, Cultural; N-Cu, non-Cultural. Native language speakers represented as E, English; N-E, non-English. Finally, cultural diversity identified as W, Western; N-W, non-Western.

**Table 1: Course Relevance**

| Variables | Levene's Test | Testing results | Mean | SD | Null Hypothesis |
|---|---|---|---|---|---|
| Defendants country of origin | P=0.258 | $F(6, 48) = 8.214$, $p<0.001$ | In: 2.82 Ja: 2.93 Ir: 2.23 Ta: 2.98 Ma: 2.40 | 1.214 | Rejected |
| Cultural diversity | P=0.243 | $t(52) = -1.877$, $p=.255$ | W: 2.68 N-W: 3.61 | 2.421 | Failed to be rejected |
| Religion and Philosophy | P=0.433 | $t(52) = -2.52$, $p=0.354$ | Ch: 1.89 N-Ch: 2.21 | 1.869 | Failed to be rejected |
| Art and Music | P= 0.589 | $t(52) = 3.54$, $p<0.001$ | Cu: 2.01 N-Cu: 3.88 | 2.001 | Rejected |
| Native language speaker | P= 0.617 | $t(52) = 4.58$, $p<.001$ | E: 3.20 N-E: 2.86 | 1.998 | Rejected |

**Table 2: Course Organization**

| Variables | Levene's Test | Testing results | Mean | SD | Null Hypothesis |
|---|---|---|---|---|---|
| Defendants country of origin | P=0.351 | $F(6, 48) = 0.766$, $p=0.453$ | In: 2.99 Ja: 3.12 Ir: 2.77 Ta: 3.45 Ma: 3.86 | 1.466 | Failed to be Rejected |
| Cultural diversity | P=0.186 | $t(52) = -0.887$ $p=.487$ | W: 3.40 N-W: 3.01 | 1.847 | Failed to be rejected |
| Religion and Philosophy | P=0.774 | $t(52) = -0.213$, $p=0.284$ | Ch: 2.55 N-Ch: 2.11 | 1.485 | Failed to be rejected |
| Art and Music | P= 0.665 | $t(52) = 0.854$, $p=0.351$ | Cu: 1.99 N-Cu: 2.88 | 1.774 | Failed to be rejected |
| Native language speaker | P= 0.411 | $t(52) = 0.889$, $p=0.661$ | E: 2.89 N-E: 3.08 | 1.965 | Failed to be rejected |

**Table 3: Course Effectiveness**

| Variables | Levene's Test | Testing results | Mean | SD | Null Hypothesis |
|---|---|---|---|---|---|
| Defendants country of origin | P=0.255 | $F(6, 48) = 0.588$, $p=0.459$ | In: 3.49 Ja: 3.85 Ir: 2.47 Ta: 2.95 Ma: 3.26 | 1.998 | Failed to be Rejected |
| Cultural diversity | P=0.655 | $t(52) = -0.887$ $p=.287$ | W: 2.99 N-W: 3.91 | 1.015 | Failed to be rejected |
| Religion and Philosophy | P=0.774 | $t(52) = -0.613$, $p=0.984$ | Ch: 3.85 N-Ch: 2.81 | 1.245 | Failed to be rejected |
| Art and Music | P= 0.886 | $t(52) = 0.334$, $p=0.851$ | Cu: 3.99 N-Cu: 3.88 | 1.377 | Failed to be rejected |
| Native language speaker | P= 0.511 | $t(52) = 0.821$, $p=0.461$ | E: 2.10 N-E: 2.48 | 1.695 | Failed to be rejected |

**Table 4: Course Interactivity**

| Variables | Levene's Test | Testing results | Mean | SD | Null Hypothesis |
|---|---|---|---|---|---|
| Defendants country of origin | P=0.114 | $F(6, 48) = 0.912$, $p=0.629$ | In: 2.47 Ja: 3.15 Ir: 3.77 Ta: 3.85 Ma: 3.06 | 2.001 | Failed to be Rejected |
| Cultural diversity | P=0.922 | $t(52) = -0.786$ $p=.445$ | W: 2.19 N-W: 3.01 | 2.655 | Failed to be rejected |
| Religion and Philosophy | P=0.877 | $t(52) = -0.243$, $p=0.784$ | Ch: 2.85 N-Ch: 2.71 | 1.895 | Failed to be rejected |
| Art and Music | P= 0.686 | $t(52) = 0.154$, $p=0.955$ | Cu: 3.74 N-Cu: 3.18 | 1.687 | Failed to be rejected |
| Native language speaker | P= 0.555 | $t(52) = 0.221$, $p=0.361$ | E: 2.80 N-E: 2.98 | 1.615 | Failed to be rejected |

**Table 5: Course Accuracy**

| Variables | Levene's Test | Testing results | Mean | SD | Null Hypothesis |
|---|---|---|---|---|---|
| Defendants country of origin | P=0.884 | $F(6, 48) = 0.382$, $p=0.229$ | In: 2.97 Ja: 2.85 Ir: 3.07 Ta: 3.15 Ma: 3.46 | 1.001 | Failed to be Rejected |
| Cultural diversity | P=0.672 | $t(52) = -0.489$ $p=.775$ | W: 3.69 N-W: 3.81 | 1.622 | Failed to be rejected |
| Religion and Philosophy | P=0.612 | $t(52) = -0.344$, $p=0.282$ | Ch: 2.81 N-Ch: 3.76 | 1.495 | Failed to be rejected |
| Art and Music | P= 0.488 | $t(52) = 0.315$, $p=0.452$ | Cu: 3.04 N-Cu: 3.78 | 1.877 | Failed to be rejected |
| Native language speaker | P= 0.510 | $t(52) = 0.321$, $p=0.391$ | E: 2.690 N-E: 2.78 | 1.156 | Failed to be rejected |

**Table 6:** Course Productivity

| Variables | Levene's Test | Testing results | Mean | SD | Null Hypothesis |
|---|---|---|---|---|---|
| Defendants country of origin | P=0.445 | F(6, 48) = 0.382, p=0.229 | In: 3.07 Ja: 2.75 Ir: 3.99 Ta: 3.45 Ma: 3.06 | 2.501 | Failed to be Rejected |
| Cultural diversity | P=0.377 | t(52) = -0.489 p=.775 | W: 2.69 N-W: 2.15 | 1.912 | Failed to be rejected |
| Religion and Philosophy | P=0.662 | t(52) = -0.344, p=0.282 | Ch: 3.71 N-Ch: 3.06 | 1.295 | Failed to be rejected |
| Art and Music | P= 0.988 | t(52) = 0.315, p=0.452 | Cu: 3.04 N-Cu: 3.98 | 1.467 | Failed to be rejected |
| Native language speaker | P= 0.980 | t(52) = 0.321, p=0.391 | E: 2.90 N-E: 2.48 | 1.311 | Failed to be rejected |

In context of course relevant, with the defendant's country of origin results from postulate 1 showed that learners (controllers) from Taiwan (Ta) have a viewpoint about impartial e-learning environments, whereas learners from India (In), Japan (Ja), Iran (Ir) and Malaysia (Ma) has Impartial to partial (positive) viewpoint about the same content. For cultural diversity, the results from postulate 2 showed that both Western and non-Western learners shared an impartial to negative cognizance about content matter. In Religion and Philosophy results from postulate 3 showed that Christian and non-Christian shared an impartial to negative cognizance about content matter. In Art and music results from postulate 4 showed that the cultural and non-Cultural showed that non-Cultural learner had positive viewpoint, Cultural had a negative viewpoint about course relevance. For native language speaker, the results from postulate 5 showed that English and non-English speakers had positive viewpoint, while non-English speaker had a negative viewpoint about course relevance.

The inconsistency obtained from postulates 1, 2 and 3 validated the theory given by Hannon and D'Netto (2007) according to them learners from different cultural backgrounds did not experience learning environment as culturally comprehensive regarding association with content. Results of postulates 4 and 5 validated the assumptions made by Hannon and Netto regarding the art & music and native language background of learners which stated that learners from different language background and cultural activities to respond differently to clamant built in e-learning course. The final response from learners on course relevance that e-learning content and training based on it was relevant to their performance as station traffic controllers and required less effort and motivation for learning an e-course.

Regarding course organization, effectiveness, interactivity, accuracy and productivity learners from different independent groups shared the same viewpoint on e-learning based environment. According to results perception, all defendants had a positive to impartial cognizance about course organization, its effectiveness, interactivity, accuracy and productivity in e-learning courses. Proper organization, of course content, interaction between participants using navigational tools, integrity of courses and content delivery, which lead to accuracy, effectiveness of course learning modules and overall how productive these courses for learners positively influence the vibe and learners cognizance of e-learning. These results denied the idea proposed by Hannon and Netto in context, of course organization, effectiveness, interactivity, accuracy and productivity.

For course organization they stated that learners from different background or origin had a different way to respond the modules built or their course organization in e-learning environment. Our research determined that railway authorities had designed, highly synchronous, structured, adaptive and flexible course with the help of navigational tools and taken strong positive feedback from learners.

For course interactivity they have stated that due to various usages of e-learning tools, their importance in some modules got differed significantly which leads to lack of interactivity. Our research determined that the railway authorities must develop advanced, adapted, organized course learning modules which utilized tools effectively and enhanced interactivity.

For course accuracy they predicted that due to lack of synchronization of courses and their interaction with others it affects the accuracy of learner. Our research concluded that railway system must use high tolerance and error free based navigational and graphical tools to improve the integrity or accuracy of course which helped learner to increase self –motivation.

For course effectiveness it directly relied on course accuracy, more accurate the course will be then more its effectiveness. Railway officials measured the effectiveness of courses based on the survey conducted by them for their learners which considered as feedback and it improves learner learning rate.

For a course productivity number of learners should be trained by a railway system to enhance their productivity within the proposed deadline.

## 5. CONCLUSION

This study provides an investigation of the Railway Station Controller based on global environments for the e - learning domain. The railway system relied heavily on technology and created a platform which just acted as learning repository for answering the problems faced by station traffic controller based on all assumptions and results. They did not recognize the learners need and identified the problem faced by learner during the analysis, design and their instructional phase. Their major motive could improve learner training by incorporating the cultural aspects of them into e-learning environment. This lack of attention with low effective training with inefficient performance metric leads to learner's negative cognizance towards e-learning.

In future we can emphasize on m-learning for station controllers. The emerging Internet of Things (IoT) and m-learning bring new opportunity for station controllers.